\documentclass[prl,twocolumn,showpacs,superscriptaddress,floatfix]{revtex4}
\usepackage{epsfig,graphicx,amsfonts}
\begin{document}

\title{On-Chip Microwave Fock States and Quantum Homodyne Measurements}

\author{M.~Mariantoni} \email{Matteo.Mariantoni@wmi.badw.de}

\affiliation{Walther-Meissner-Institut, Bayerische Akademie der
Wissenschaften, Walther-Meissner-Strasse 8, D-85748 Garching,
Germany}

\author{M.J.~Storcz}

\affiliation{Department Physik, CeNS and ASC, LMU,
Theresienstrasse 37, D-80333 M\"unchen, Germany}

\author{F.K.~Wilhelm}

\affiliation{Department Physik, CeNS and ASC, LMU,
Theresienstrasse 37, D-80333 M\"unchen, Germany}

\author{W.D.~Oliver}

\affiliation{MIT Lincoln Laboratory, 244 Wood Street, Lexington,
Massachussets 02420, USA}

\author{A.~Emmert}

\affiliation{Walther-Meissner-Institut, Bayerische Akademie der
Wissenschaften, Walther-Meissner-Strasse 8, D-85748 Garching,
Germany}

\author{A.~Marx}

\affiliation{Walther-Meissner-Institut, Bayerische Akademie der
Wissenschaften, Walther-Meissner-Strasse 8, D-85748 Garching,
Germany}

\author{R.~Gross}

\affiliation{Walther-Meissner-Institut, Bayerische Akademie der
Wissenschaften, Walther-Meissner-Strasse 8, D-85748 Garching,
Germany}

\author{H.~Christ}

\affiliation{Max-Planck Institute for Quantum Optics,
Hans-Kopfermann-Strasse 1, D-85748 Garching, Germany}

\author{E.~Solano}

\affiliation{Max-Planck Institute for Quantum Optics,
Hans-Kopfermann-Strasse 1, D-85748 Garching, Germany}

\affiliation{Secci\'{o}n F\'{\i}sica, Departamento de Ciencias,
Pontificia Universidad Cat\'{o}lica del Per\'{u}, Apartado 1761,
Lima, Peru}

\begin{abstract}
We propose a method to couple metastable flux-based qubits to
superconductive resonators based on a quantum-optical Raman
excitation scheme that allows for the deterministic generation of
stationary and propagating microwave Fock states and other weak
quantum fields. Moreover, we introduce a suitable microwave
quantum homodyne technique, with no optical counterpart, that
enables the measurement of relevant field observables, even in the
presence of noisy amplification~devices.
\end{abstract}


\maketitle

A two-level atom coupled to a single mode of a quantized
electromagnetic field is arguably the most fundamental system to
exhibit matter-radiation interplay. Their interaction, described
by the Jaynes-Cummings (JC) model, arises naturally in the realm
of cavity quantum electrodynamics (CQED) in the
microwave~\cite{Raimond:2001:a} and optical
domains~\cite{Mabuchi:2002:a}. There, a variety of nonclassical
states (e.g., squeezed, Schr\"odinger cat, and Fock states) and
other remarkable phenomena and applications (e.g., entanglement
and elements of quantum logic~\cite{Nielsen:2000:a}) have been
proposed and realized. Other physical systems, like trapped
ions~\cite{Leibfried:2003:a}, can reproduce the JC dynamics and,
consequently, exploit this analogy for similar purposes. The
intracavity field in CQED and the motional field of the trapped
ions are typically detected through a suitable transfer of
information to measurable atomic degrees
of~freedom~\cite{Raimond:2001:a,Leibfried:2003:a}. In the case of
propagating fields, quantum homodyne
detection~\cite{Leonhardt:1997:a}, a method based on efficient
photodetectors, leads to
quantum~state~tomography~\cite{Lvovsky:2001:a}.

Given its relevance for quantum communication, single-photon
sources have been pursued in the optical
domain~\cite{Kuhn:2002:a,Keller:2004:a}. Recently, several
CQED-related experiments have been performed in tunable,
solid-state systems. Quantum dots in photonic band-gap structures
have been used as single photon sources~\cite{Badolato:2005:a} and
superconducting qubits have been coupled to on-chip
cavities~\cite{Wallraff:2004:a,Chiorescu:2004:a}. In addition,
microwave squeezing with Josephson parametric
amplifiers~\cite{Movshovich:1990:a} and aspects of the
quantum-statistical nature of GHz photons in mesoscopic conductors
have been demonstrated~\cite{Gabelli:2004:a}.

In this Letter, we propose the implementation of a deterministic
source of microwave Fock states, or linear superpositions of them,
at the output of a superconducting resonator containing a flux
qubit. A Raman-like scheme~\cite{Francsa:Santos:2001:a} determines
the coupling between the cavity and the qubit that consists of two
metastable states of a three-level system in a
$\Lambda$-type~configuration~\cite{Yang:2004:a,Murali:2004:a,Liu:2004:a,Siewert:2005:a}.
Furthermore, we show that relevant observables of these weak
quantum signals can be characterized by a \emph{microwave quantum
homodyne measurement}~(MQHM) scheme. This is a specially
engineered technique based on a hybrid
ring~junction~\cite{Collin:2000:a,Sarovar:2005:a} acting as an
on-chip \textit{\mbox{microwave} beam splitter}~(MBS). Given that
direct microwave photodetectors are not yet available, the
measurement device consists of linear phase-insensitive amplifiers
\mbox{followed} by square-law~detectors~(and/or mixers)
\mbox{and}~an~oscilloscope~(or spectrum analyzer). These
fundamental tools represent building blocks to establish on-chip
quantum information transfer~between~qubits.

A prototypical example of a flux-based quantum circuit is the
radio-frequency~(RF) superconducting quantum-interference
device~(SQUID), a superconducting loop interrupted by a single
Josephson tunnel junction. The RF SQUID Hamiltonian is
\begin{equation}
\hat{H}^{}_{\rm S} \ = \ \frac{\hat{Q}^2}{2 C_{\rm j}} +
\frac{(\hat{\Phi} - \Phi_{\rm x})^2}{2 L_{\rm s}} - E_{\rm J} \cos
\bigg( 2 \pi \frac{\hat{\Phi}}{\Phi_0} \bigg) , \label{H:S}
\end{equation}
where $\hat{Q}$ is the charge stored on the junction capacitor
$C_{\rm j}$, $\hat{\Phi}$ is the total flux threading the loop
(with $[ \hat{\Phi} , \hat{Q} ] = i \hbar$), $\Phi_{\rm x}$ is an
externally applied quasi-static flux bias, $L_{\rm s}$ is the
self-inductance of the loop, $E_{\rm J} \equiv I_{{\rm c} 0}
\Phi_0 / 2 \pi$ is the Josephson coupling energy, $I_{{\rm c} 0}$
is the junction critical current, and $\Phi_0 = h / 2 e$ is the
flux quantum.
\begin{figure}[t]
\centering{
\includegraphics[width=0.99\columnwidth,clip=]{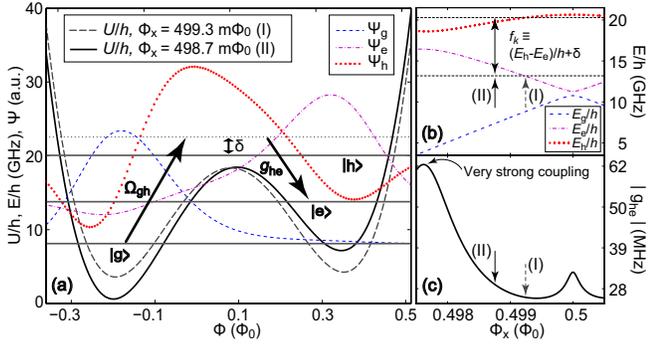}}
\caption{(Color online)~(a)~Potential profile of the RF SQUID for
two distinct values of $\Phi_{\rm x}$~[(I) and (II)] and wave
functions for the first three lowest energy levels ($| {\rm g}
\rangle$, $| {\rm e} \rangle$, and $| {\rm h} \rangle$).
Parameters are given in Table I. The Raman scheme is indicated
with arrows. (b)~RF SQUID energy-band diagram near $\Phi_0 / 2$
plotted vs.~$\Phi_{\rm x}$. Zero detuning case, point (I), and
large detuning case, point (II). (c)~Absolute value of the vacuum
Rabi frequency $g_{\rm he}$ as a function of $\Phi_{\rm x}$. The
coupling reaches 56~MHz at the anticrossing between levels $| {\rm
e} \rangle$ and $| {\rm h} \rangle$.}
\label{Figure:1:abc:MMariantoni}
\end{figure}
For appropriate design parameters, and close to half-integer
values of $\Phi_{\rm x} / \Phi_0$, the RF SQUID potential profile
becomes a relatively shallow double well whose asymmetry can be
tuned by setting $\Phi_{\rm x}$~(see
Fig.~\ref{Figure:1:abc:MMariantoni}). In this case, the two lowest
eigenstates~$| {\rm g} \rangle$ and $| {\rm e} \rangle$ are
metastable and localized in the left and right wells respectively,
whereas the excited state~$| {\rm h} \rangle$ is delocalized with
energy above the barrier. As will be shown later, these levels are
suitable for implementing a Raman excitation scheme. The energy
levels can be tuned by statically biasing $\Phi_{\rm x}$ during
the experiment and transitions between states are driven by pulsed
AC excitations. Because of their relatively large
excited-state~lifetimes, persistent-current (PC)
qubits~\cite{Chiorescu:2004:a} can also be considered for the
purposes of this work.

\begin{figure}[t]
\centering{%
    \includegraphics[width=0.99\columnwidth,clip=]{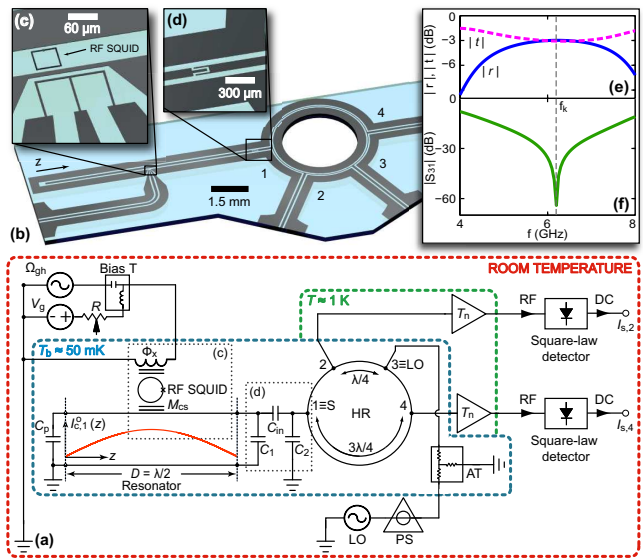}}
\caption{(Color online)~(a)~Sketch of the entire generation and
measurement network ($C_{\rm in}$, $C_1$, and $C_2$: capacitive
$\Pi$ network representing the resonator input port acting as a
mirror; $C_{\rm p}$: parasitic capacitor at the cavity
open-circuit ending; HR: hybrid ring; PS: phase shifter; AT:
attenuator). (b) Asymmetric CWG resonator with integrated HR. (c)
On-chip antenna providing the classical driving $\Omega_{\rm gh}$.
(d) Resonator-CWG coupling region. (e) The waves traveling around
the HR interfere resulting in the plotted reflection and
transmission amplitude patterns. (f) Isolation between ports three
and one of the~HR.} \label{Figure:2:abcdef:MMariantoni}
\end{figure}

The segment of superconducting coplanar waveguide (CWG) shown in
Figs.~\ref{Figure:2:abcdef:MMariantoni}~(a),~(b),~and~(d) is one
realization of an open-circuited transmission line resonator
capacitively coupled to a CWG transmission line. Such a cavity is
characterized by eigenenergies with transition angular frequencies
$\omega_k = 2 \pi f_k$ that are much larger than the thermal
energy at cryogenic temperatures. Its Hamiltonian is
\mbox{$\hat{H}^{}_{\rm C} = \sum_k \hbar \omega_k \Big(
\hat{a}^{\dagger}_k \hat{a}^{}_k + 1/2 \Big)$, where
$\hat{a}^{\dagger}_k$} and $\hat{a}^{}_k$ are the bosonic creation
and annihilation operators for mode $k$. Current and voltage,
corresponding to magnetic and electric fields respectively, are
conjugate operators associated with the quantized resonator,
$\hat{I}^{}_{\rm c} ( z , t ) = ( \partial /
\partial t )\,\hat{\vartheta} ( z , t )$, where $z$ is the spatial coordinate for the
superconducting inner strip and $\hat{\vartheta} ( z , t )$ is the
normal mode expansion of the cavity field~\cite{Blais:2004:a}. The
vacuum current for the odd, $k_{\rm o}$, and even, $k_{\rm e}$,
cavity modes is $I^{0}_{{\rm c} , k_{\rm o / e}} ( z ) =
\sum_{k_{\rm o} = 1}^{\tilde{k}_{\rm o}} \sqrt{\hbar
\omega_{k_{\rm o}} / ( D l )} \cos \big( k_{\rm o} \pi z / D \big)
+ \sum_{k_{\rm e} = 2}^{\tilde{k}_{\rm e}} \sqrt{\hbar
\omega_{k_{\rm e}} / ( D l )} \sin \big( k_{\rm e} \pi z / D
\big)$, where $\tilde{k}_{\rm o}$, $\tilde{k}_{\rm e}$ are upper
cutoffs for the odd and even modes
respectively~\cite{Blais:2004:a}, $D = \lambda / 2$ is the length
of the resonator ($\lambda$ is the full wavelength corresponding
to $\omega_{k_{\rm o} = 1}$), and $l$ is its total series
inductance per unit length. Hereafter, the cavity is chosen to be
operated at $k_{\rm 0} = 1$ and is assumed to have an external
quality factor $Q_{\rm x} \sim 10^4$ at $f_1 = \omega_1 / 2 \pi
\sim 10$~GHz, corresponding to a cavity decay rate $\kappa_{\rm c}
\sim 1$~MHz~\cite{Wallraff:2004:a}. Hence, at a base temperature
$T_{\rm b} \sim 50$~mK, the mean number of thermal photons is
$\langle n_{\rm th} \rangle = [ \exp ( \hbar \omega_1 / k_{\rm B}
T_{\rm b} ) - 1 ]^{-1} \approx~10^{-4}$ and the cavity mode can be
considered to be in the vacuum state $| 0 \rangle$.

Embedding the RF SQUID in the CWG cavity allows an inductive
coupling between any two levels of the RF SQUID and a single
cavity mode $k$. The resulting interaction Hamiltonian is
\mbox{$\hat{H}^{}_{\rm I} = - M^{}_{\rm cs} \hat{I}^{}_{{\rm c} ,
k} \hat{I}^{}_{\rm s}$}, where $\hat{I}_{{\rm c} , k}$ and
$\hat{I}^{}_{\rm s}$ are the resonator and the RF SQUID current
operators respectively and $M_{\rm cs}$ is their mutual
inductance. Using the explicit forms of the current operators, we
find
\begin{equation}
\hat{H}^{}_{\rm I}\,=\,- \ ( M_{\rm cs} / L_{\rm s} )\,I^{0}_{{\rm
c} , k} ( z ) \ ( \hat{\Phi} - \Phi_{\rm x} ) \ i \,[
\hat{a}^{\dagger}_k ( t ) - \hat{a}^{}_k ( t ) ] . \label{H:I}
\end{equation}
The RF SQUID can be positioned near one of the \mbox{antinodes} of
the vacuum current [e.g., see
Figs.~\ref{Figure:2:abcdef:MMariantoni}~(a)~and~(b) for $k_{\rm o}
= 1$] and can be biased to yield maximum coupling for any two of
its eigenstates $| i \rangle$ and $| j \rangle$. The interaction
matrix element between these levels represents their coupling
strength with mode $k$ and it is used to define the vacuum Rabi
frequency $g_{ij} = -\,( M_{\rm cs} / L_{\rm s} ) I^{0}_{{\rm c} ,
k} ( z ) \langle i | \hat{\Phi} | j \rangle / h$. Moreover, when
operating the system in the dispersive regime, the corresponding
RF SQUID decay rates~$\gamma^{}_{ij}$ become $\gamma^{\rm
eff}_{ij} = \gamma^{}_{ij} ( g_{ij} / \delta_{ij} )^2$, where
$\delta_{ij}$ is the detuning between mode $k$ and the transition
under consideration. The coupling $g_{\rm he}$, the effective
decay rate $\gamma^{\rm eff}_{\rm he}$, and other relevant
quantities have been calculated for both RF SQUIDs and PC qubits
and typical results are~reported~in~Table~I.

A main application of the system illustrated above is the
generation of single photons at frequency $f_{k}$ in a manner
similar to a quantum-optical Raman
scheme~\cite{Francsa:Santos:2001:a,Keller:2004:a}. After preparing
the RF SQUID in level $| {\rm g} \rangle$, the transition $| {\rm
g} \rangle \leftrightarrow | {\rm h} \rangle$ is driven by a
classical excitation with Rabi frequency $\Omega_{\rm gh} \sim
g_{\rm he}$ and detuned by the amount $\delta$. The same
transition is detuned from the resonator mode $k$ by an amount
$\Delta \gg \delta$, resulting in a comparatively negligible
coupling. On the other hand, the $| {\rm h} \rangle
\leftrightarrow | {\rm e} \rangle$ transition is the only one
coupled to mode $k$ and it is also detuned by $\delta$ [see
Figs.~\ref{Figure:1:abc:MMariantoni}~(a) and~(b)]. Choosing
$\delta \gg \max \left[ \Omega_{\rm gh} , g_{\rm he} \right]$,
level $| {\rm h} \rangle$ can be adiabatically
eliminated~\cite{Francsa:Santos:2001:a,Keller:2004:a}, thus
leading to the effective second-order~Hamiltonian
\begin{eqnarray}
\hat{H}^{}_{{\rm eff}} & = & \hbar \frac{\Omega_{{\rm
gh}}^2}{\delta} | {\rm g} \rangle \langle {\rm g} | + \hbar
\frac{g_{{\rm he}}^2}{\delta} | {\rm e} \rangle \langle
{\rm e} | \hat{a}^{\dagger}_k \hat{a}^{}_k + {} \nonumber\\
& & + \ \hbar g_{{\rm eff}} \left( | {\rm g} \rangle \langle {\rm
e} | \hat{a}^{}_k + | {\rm e} \rangle \langle {\rm g} |
\hat{a}^{\dagger}_k \right) , \label{H:eff}
\end{eqnarray}
where $g_{\rm eff} = ( \Omega_{\rm gh} / \delta ) g_{\rm he}$ is
the effective Raman coupling. The first two terms at the r.h.s. of
Eq.~(\ref{H:eff}) are AC Zeeman shifts, while the last term
describes an effective anti-JC dynamics, inducing transitions
within the $\{ | {\rm g} \rangle | n \rangle , | {\rm e} \rangle |
n + 1 \rangle \}$ subspaces. The AC Zeeman shifts associated with
the transition of interest~$\{ | {\rm g} \rangle | 0 \rangle , |
{\rm e} \rangle | 1 \rangle \}$, can be compensated by retuning
the classical driving frequency. When the strong-coupling regime
is reached, $g_{\rm eff} \gtrsim \max \left[ \kappa_{\rm c} ,
\gamma^{\rm eff}_{\rm he} \right]$, an effective $\pi$-pulse
realizes a complete transfer of population from state $| {\rm g}
\rangle | 0 \rangle$ to state $| {\rm e} \rangle | 1 \rangle$.
This process leads to the creation of a microwave Fock state $| 1
\rangle$ inside the resonator that will leak out in a time $\sim 1
/ \kappa_{\rm c}$. In the case of weak-coupling, the photon leaks
to the outer world as soon as it is generated inside the cavity,
thereby realizing a deterministic single-photon source. Tailoring
the photon pulse shape would require a time-dependent
classical~driving~$\Omega_{\rm gh} ( t )$~\cite{Keller:2004:a}.

If the initial qubit-cavity state is $( \cos \theta | {\rm g}
\rangle + e^{i \phi} \sin \theta | {\rm e} \rangle ) | 0 \rangle$,
then the above Raman $\pi$-pulse would map it onto the cavity
field ($| {\rm e} \rangle | 0 \rangle$ is a dark state of the
anti-JC dynamics). In this way, we would be able to produce an
outgoing field state $\cos \theta | 1 \rangle + e^{i \phi} \sin
\theta | 0 \rangle$.

The proposed Raman scheme allows for generating single photons
without initialization of the qubit in the excited state $| {\rm
e} \rangle$~($| {\rm g} \rangle | 0 \rangle$ is not a dark state
of the anti-JC dynamics). Raman pulses are fast compare to STIRAP
(adiabatic) techniques~\cite{Kuhn:2002:a} and exploit the well
defined qubit-cavity coupling~\cite{Keller:2004:a}, intrinsic in
our scheme.

\begin{table}[b]
\caption{Typical parameters and relevant quantities for RF SQUIDs
and PC qubits, coupled to a
50~$\Omega$~resonator~\cite{Resonator:extra:parameters}.}
\begin{ruledtabular}
   \begin{tabular}{c c c c c c c c c c c}
      & $I_{{\rm c} 0}$ & $C_{\rm j}$ & $L_{\rm s}$ & $M_{\rm cs}$ & $\frac{\Delta E_{\rm he}}{h}$
      & $I^{0}_{{\rm c} , 1}$ & $| g_{\rm he} |$ & $\gamma^{\rm eff}_{\rm he}$
   \\
      & \footnotesize{($\mu$A)} & \footnotesize{(fF)} & \footnotesize{(pH)} & \footnotesize{(pH)}
      & \footnotesize{(GHz)} & \footnotesize{(nA)} & \footnotesize{(MHz)} & \footnotesize{(kHz)}
   \\
   \hline
      RF SQUID & 1.4 & 100 & 266 & 22.2 & 6.2 & 32 & 28 & 2.5 \\
      PC qubit & 0.64 & 7 & 17 & 1.4 & 5.8 & 31.5 & 30 & 2.5
   \end{tabular}
\end{ruledtabular}
   \label{Table:1:MMariantoni}
\end{table}

Measurement schemes based on classical
homodyning~\cite{Collin:2000:a} are insufficient to resolve
nonclassical field states. On the other hand, microwave
single-photon detectors in mesoscopic systems do not exist to our
knowledge. Here, we propose an MQHM technique as a means to
measure relevant observables of weak quantum signals, even at the
level of single photons. This can be implemented in three main
steps. First, a signal (S) and a local oscillator (LO),
characterized by the same angular frequency $\omega_{\rm S} =
\omega_{\rm LO}$, are coherently superposed at a suitably designed
MBS cooled to base
temperature~[Fig.~\ref{Figure:2:abcdef:MMariantoni}~(a)]. Second,
the microwave fields at the MBS output ports are amplified at low
temperatures by means of linear phase-insensitive amplifiers.
Third, the amplified signals are then downconverted to DC currents
$I_{{\rm s} , 2 / 4}$, proportional to the energies of the input
signals, via square-law~detectors~(and/or mixers) at room
temperature~\cite{Collin:2000:a}. The DC currents can be measured
as voltages with an oscilloscope. Adequate manipulation of these
measurements will lead to signal information
with~minimal~noise~background.

The MBS is realized using a suitable four-port device: the
\mbox{hybrid ring}~[see
Figs.~\ref{Figure:2:abcdef:MMariantoni}~(a)~and~(b)]. The
advantageous coplanar design proposed here can easily be scaled
and preferably fabricated with low resistivity conductors. We now
extend the classical theory of hybrid rings in
Ref.~\cite{Collin:2000:a} to the quantum regime by analogy with an
optical beam splitter. With only the vacuum incident at ports two
and four, and up to a global phase common to both input beams, the
reduced quantum input-output relations of a~lossless~MBS~are
\begin{eqnarray}
\left[ \begin{array}{c}
\hat{a}^{}_{2} \\
\hat{a}^{}_{4}
\end{array} \right] \ = \ \left[
\begin{array}{cc}
r & t \\
- t^{*} & r^{*}
\end{array} \right] \
\left[ \begin{array}{c}
\hat{a}_{\rm S} \\
\hat{a}_{\rm LO}
\end{array} \right] . \label{HR:in:out:q}
\end{eqnarray}
Here, $r$ and $t$ are the complex, frequency-dependent reflection
and transmission coefficients, $\hat{a}^{}_{\rm S}$ and
$\hat{a}^{}_{\rm LO}$ are the signal and LO port operators
respectively. The latter is chosen to be a classical coherent
field that is characterized by its complex amplitude
\mbox{$\alpha^{}_{\rm LO} = | \alpha^{}_{\rm LO} | \exp ( i
\theta^{}_{\rm r} )$}, where $| \alpha^{}_{\rm LO} |$ is the norm
of the field amplitude and $\theta^{}_{\rm r}$ is its relative
phase with respect to S. The numerical simulations plotted in
Fig.~\ref{Figure:2:abcdef:MMariantoni}~(e) show that the MBS can
be {\it balanced} over a broad bandwidth around the desired
operation frequency $f_k$, i.e.,~$r = t = 1 / \sqrt{2}$~($-$3~dB),
with outputs $\hat{a}^{}_2 = ( \hat{a}^{}_{\rm S} +
\hat{a}^{}_{\rm LO}) / \sqrt{2}$ and $\hat{a}^{}_4 = ( -
\hat{a}^{}_{\rm S} + \hat{a}^{}_{\rm LO}) / \sqrt{2}$.

At this point, standard optical quantum homodyning would require
the use of photodetectors at each output port of the MBS. This
would allow different measurements of photocurrents $I_{{\rm p} ,
2 / 4}$, proportional to different realizations $N$ of the
observable $\hat{n} = \hat{a}^{\dagger} \hat{a}$. By computing the
difference $I_{{\rm p} , 2} - I_{{\rm p} , 4} \propto N_2 - N_4 =
2 | \alpha_{\rm LO} | X_{\theta_{\rm r}}$, where the LO operators
were replaced by their complex amplitudes, realizations of the
quadrature $\hat{X}_{\theta_{\rm r}} \equiv (
\hat{a}^{\dagger}_{\rm S} e^{ i \theta_{\rm r}} + \hat{a}^{}_{\rm
S} e^{ - i \theta_{\rm r}} ) / 2$, enhanced by a factor $2 |
\alpha_{\rm LO} |$, could be obtained. With the complete histogram
of these measurements, all moment averages $\langle
\hat{X}^{p}_{\theta_{\rm r}} \rangle$, $\forall p \in \mathbb{N}$,
and full reconstruction of the associated Wigner function via
quantum tomography might be evaluated~\cite{Leonhardt:1997:a}. In
absence of microwave photodetectors, the MBS output signals must
go through linear amplifiers, square-law detectors and/or mixers
before being measured at the oscilloscope. These conditions impose
severe restrictions in the measurement process with no counterpart
in the optical regime and require additional
theoretical~considerations.

The output signals, referred to the input, of linear
phase-insensitive amplifiers can be written as ${\hat X}_{{\tilde
a}_2} = {\hat X}_{a_2} + {\hat X}_{\xi_2}$ and ${\hat X}_{{\tilde
a}_4} = {\hat X}_{a_4} + {\hat X}_{\xi_4}$, where ${\hat X}_a
\equiv ( {\hat a} + {\hat a}^{\dagger} ) / 2$ and ${\hat X}_{\xi}
\equiv ( {\hat \xi} + {\hat \xi}^{\dagger} ) /
2$~\cite{Caves:1982:a}. The added noises at each arm, $\hat{ \xi }
_{2 / 4}$, are random, uncorrelated, and characterized by (almost)
the same noise temperature $T_{\rm n}$, leading to a mean photon
number $\langle {\hat n}_{\xi} \rangle = \langle {\hat
\xi}^{\dagger} {\hat \xi} \rangle = k_{\rm B} T_{\rm n} / \hbar
\omega_k$. The difference between the measured currents produced
by the square-law detectors, $I_{{\rm s} , 2} - I_{{\rm s} , 4}$,
is proportional to different realizations of the following
observable
\begin{eqnarray}
\!\!\!\!\!\! {\hat {\tilde a}}^{\dagger}_2 {\hat {\tilde a}}^{}_2
- {\hat {\tilde a}}^{\dagger}_4 {\hat {\tilde a}}^{}_4 =
\!\!\!\!\! & & {\hat a}^{\dagger}_{\rm S} {\hat a}^{}_{\rm LO} +
{\hat a}^{\dagger}_{\rm LO} {\hat a}^{}_{\rm S} + {\hat
\xi}_{2}^{\dagger} {\hat \xi}^{}_2 - {\hat \xi}_{4}^{\dagger}
{\hat \xi}^{}_4 \nonumber\\
\!\! & & + \ {\hat \xi}_{2}^{\dagger} \frac{ ( {\hat a}^{}_{\rm S}
+ {\hat a}^{}_{\rm LO} ) }{\sqrt{2}} + {\hat \xi}^{}_2 \frac{ (
{\hat a}^{\dagger}_{\rm S} + {\hat a}^{\dagger}_{\rm LO} )
}{\sqrt{2}} \nonumber\\
\!\! & & - \ {\hat \xi}_{4}^{\dagger} \frac{ ( - {\hat a}^{}_{\rm
S} + {\hat a}^{}_{\rm LO} ) }{\sqrt{2}} - {\hat \xi}^{}_4 \frac{ (
- {\hat a}^{\dagger}_{\rm S} + {\hat a}^{\dagger}_{\rm LO} )
}{\sqrt{2}} . \label{n2:m:n4:xi}
\end{eqnarray}
Repeating the measurement~procedure, we~can~average
\begin{eqnarray}
\langle {\hat {\tilde a}}^{\dagger}_2 {\hat {\tilde a}}^{}_2 -
{\hat {\tilde a}}^{\dagger}_4 {\hat {\tilde a}}^{}_4 \rangle = 2 |
\alpha^{}_{\rm LO} | \langle \hat{X}^{}_{\theta_{\rm r}} \rangle ,
\label{n2:m:n4:mean}
\end{eqnarray}
given that $\langle {\hat \xi_2} \rangle = \langle {\hat \xi_4}
\rangle = 0$ and the reasonable assumption $\langle {\hat
n}_{\xi_2} \rangle \sim \langle {\hat n}_{\xi_4} \rangle$.
Equation~(\ref{n2:m:n4:mean}) shows that the proposed MQHM allows
the measurement of the enhanced mean value of the quadrature
$\hat{X}_{\theta_{\rm r}}$ with \emph{negligible noise
disturbance}. This important physical quantity is sensitive to
coherence: it is zero for any Fock state $| n \rangle$ and $\cos{
\theta_{\rm r} } / 2$ for the superposition $( | 0 \rangle + | 1
\rangle ) / \sqrt{2}$. However, the method illustrated above does
not lead to a measurement of $\langle \hat{X}^{2}_{\theta_{\rm r}}
\rangle$ since the amplifier noise shadows the information
contained in the signal. Instead, we now propose to send both MBS
outputs through mixers, using the same calibrated local
oscillator, and then evaluate the ensemble~average of the measured
products
\begin{eqnarray}
\langle {\hat{X}}^{}_{\tilde{a}_2 , {\theta_{\rm r}}}
{\hat{X}}^{}_{\tilde{a}_4 , {\theta_{\rm r}}} \rangle = -
\frac{1}{2} \langle \hat{X}^{2}_{{\theta_{\rm r}}} \rangle +
\frac{1}{2} \langle \hat{X}^{2}_{\rm LO , {\theta_{\rm r}}}
\rangle . \label{n2:t:n4:mean}
\end{eqnarray}
Here, we used $\langle {\hat X}_{\xi_2} {\hat X}_{\xi_4} \rangle =
\langle {\hat X}_{\xi_2} \rangle \langle {\hat X}_{\xi_4} \rangle
= 0$ and assumed $\langle {\hat X}^{2}_{\rm LO , {\theta_{\rm r}}}
\rangle$ to be known~(after performing an adequate network
calibration) or removable (e.g., via a modulation technique).
Equation~(\ref{n2:t:n4:mean}) shows a remarkably simple way of
measuring $\langle \hat{X}^{2}_{{\theta_{\rm r}}} \rangle$ with
\emph{minimal noise disturbance}.

A precise measurement of $\langle \hat{X}_{\theta_{\rm r}}
\rangle$ and $\langle \hat{X}^{2}_{\theta_{\rm r}} \rangle$, as
shown in Eqs.~(\ref{n2:m:n4:mean})~and~(\ref{n2:t:n4:mean}),
requires random and uncorrelated noise, a sufficient number of
repetition measurements, and adequate calibration, if the
difference of $\langle {\hat n}_{\xi} \rangle$~($\langle {\hat
n}_{\xi} \rangle \sim 21$ for $T_{\rm n} \sim 10$~K) for the two
amplifiers is not sufficiently small. It is known that the
knowledge of $\langle \hat{X}_{\theta_{\rm r}} \rangle$ and
$\langle \hat{X}^{2}_{\theta_{\rm r}} \rangle$ provides complete
information about Gaussian states and a simple criterion for
discriminating Fock states. Furthermore, we conjecture here on the
possibility of measuring $\langle \hat{X}^{p}_{\theta_{\rm r}}
\rangle$, $\forall p \in \mathbb{N}$, $\forall {\theta_{\rm r}}
\in [ 0 , 2 \pi )$~[see PS in
Fig.~\ref{Figure:2:abcdef:MMariantoni}~(a)] under the conditions
described above, allowing a complete reconstruction of the Wigner
function in the microwave domain.

In conclusion, we proposed a new scheme for the deterministic
generation of intracavity and propagating microwave Fock states or
linear superpositions of them. We showed also how to realize MQHM
for measuring first and second-order field quadrature moments.
These proposals are essential tools for the implementation of
quantum-optical CQED and linear optics in the microwave domain
with superconducting devices on a~chip.

The authors thank C.~M.~Caves, S.~M.~Girvin, and D.~C.~Glattli for
useful discussions. MM and WDO would like to acknowledge
K.~R.~Brown, D.~E.~Oates, and Y.~Nakamura for fruitful discussions
and T.~P.~Orlando and M.~Gouker for their support. This work has
been partially supported by the DFG through SFB~631. ES
acknowledges EU support through the RESQ project.

\end{document}